\documentclass[twocolumn,twoside,slac_two]{revtex4}
\usepackage{graphicx}
\usepackage{fancyhdr}
\begin{document}

\title{GeV Gamma-ray Emission from the Binary PSR B1259-63/SS2883 
During the 2010 Periastron Passage}

%

\author{Masaki Mori}
\affiliation{Department of Physical Sciences, Ritsumeikan University, 
Kusatsu, 525-8577 Shiga, Japan}
\author{Akiko Kawachi}
\affiliation{Department of Physics, Tokai University, 
Hiratsuka, Kanagawa 259-1292, Japan}
\author{Shigehiro Nagataki}
\affiliation{Yukawa Institute for Theoretical Physics, Oiwake-cho, 
Kitashirakawa, Sakyo-ku, Kyoto 606-8502, Japan}
\author{Jumpei Takata}
\affiliation{Department of Physics, University of Hong Kong, 
Pokfulam Road, Hong Kong, China}

\begin{abstract}
PSR B1259-63/SS2883 is a binary system which consists of
a 48-ms radio pulsar and a massive star in a highly
eccentric orbit with a period of about 3.4 years.
Non-pulsed and non-thermal emissions from this binary
have been reported in the radio, X-ray and TeV gamma-ray
energy ranges. 
Light curves in the radio and X-ray bands showed
characteristic double-peaked features which
can be attibuted to the interactions of the pulsar
wind and the Be disk during the crossings by the
pulsar. The TeV light curves around periastron
 differ between 2004 and 2007 observations, and the feature
is not conclusive.
We report a detection of GeV gamma-ray emission
around the periastron passage in December 2010
with {\sl Fermi}-LAT.
\end{abstract}

\maketitle

\thispagestyle{fancy}


\section{INTRODUCTION}

The PSR B1259-63/SS2883 system is one of a few binary systems 
detected in TeV gamma-ray energies.
Gamma-rays should be emitted via interaction of high-speed 
wind from the 48-ms pulsar with the Be star wind and disk.
The elliptic orbit with long (3.4-yr) period offers a unique 
experimental field of wind interaction with varying distance 
between the pulsar and the Be star \cite{Kaw04, Oka11}.
In 2004 and 2007, H.E.S.S. detected TeV gamma-rays as 
a marginal pre-periastron peak and a clear post-periastron 
peak \cite{Aha05, Ace09}.
We studied the 2010 periastron (December 14) passage in 
GeV gamma-rays using the {\sl Fermi}-LAT data and compared 
the result with SPH simulation \cite{Oka11}.

\section{FERMI-LAT OBSERVATION}

\begin{figure}[b]
\centering
\includegraphics[width=75mm]{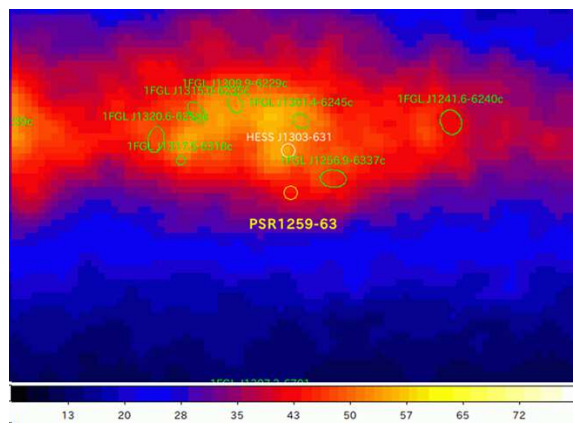}
\caption{Fermi-LAT countmap (200 MeV--10 GeV, Aug. 4, 2008 -- Feb. 9, 2011).
Gamma-ray signal from PSR 1259-63/SS2883 is not significant.} \label{fig:f1}
\end{figure}

\begin{figure}[t]
\centering
\includegraphics[width=75mm]{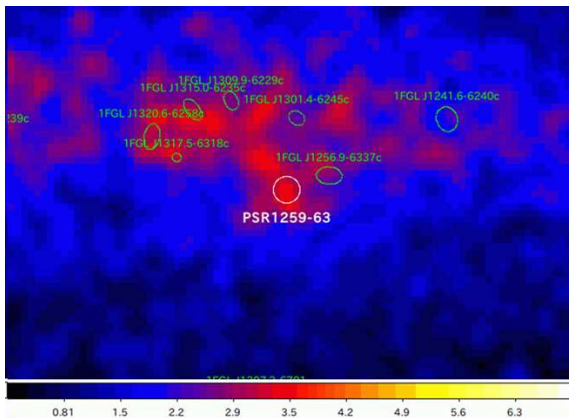}
\caption{Fermi-LAT countmap (200 MeV--10 GeV, Dec. 22, 2010 -- Jan. 21, 2011).
There is a hint of gamma-ray signal from PSR 1259-63/SS2883 with 
{\sl TS}=5.} \label{fig:f2}
\end{figure}

\begin{figure}[t]
\centering
\includegraphics[width=75mm]{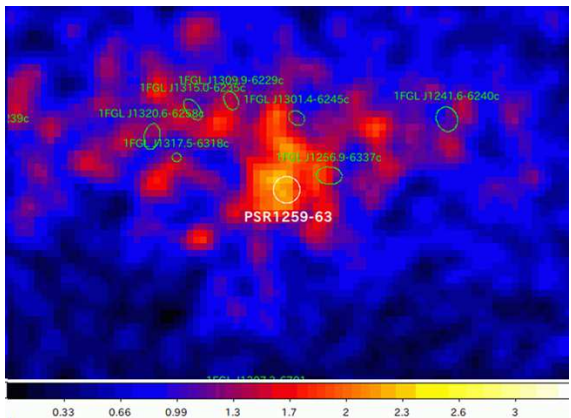}
\caption{Fermi-LAT countmap (200 MeV--10 GeV, Jan. 21, 2011 -- Feb. 9, 2011).
Gamma-ray signal from PSR 1259-63/SS2883 is significant with 
{\sl TS}=58.} \label{fig:f3}
\end{figure}

{\sl Fermi}-LAT data were obtained via Fermi Science Support Center 
and analyzed using the {\sl Fermi Science Tools} (v9r17p0) 
with P6\_V3\_DIFFUSE instrument response function
by the standard pipeline \cite{FSSC}.
Examples of countmaps are shown in Figures 1 to 3.
For the whole observation period (August 4, 2008 -- February 9, 2011)
the gamma-ray signal from PSR 1259-63/SS2883 is not signifincant
(Figure \ref{fig:f1}). Howerver, in the month after the periastron
passage (December 22, 2010 -- January 21, 2011) there is a
hint of signal (Figure \ref{fig:f2}) with a {\sl TS} (test
statistic) value of 5 (which means it is significant at $2\sigma$ level)
and in the following month (January 21, 2011
-- February 9, 2011) the gamma-ray signal is highly significant
 (Figure \ref{fig:f3}) with a {\sl TS} value of 58 ($7\sigma$ level).

\section{LIGHT CURVES}

\begin{figure}[t]
\centering
\includegraphics[width=75mm]{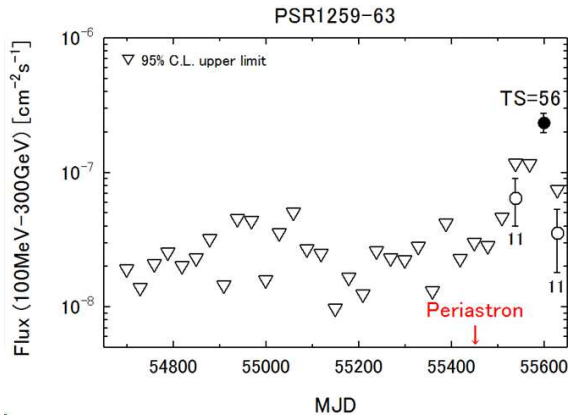}
\caption{30-day-bin light curve. Positive detection is observed only 
after the periastron. Open circles are fluxes with marginal significance 
($\sim3\sigma$).} \label{fig:f4}
\end{figure}

\begin{figure}[t]
\centering
\includegraphics[width=75mm]{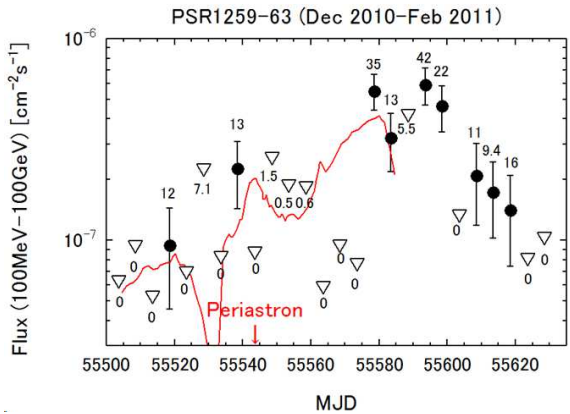}
\caption{Close-up of light curve in 5-day bins around periastron. 
Points are plotted including marginal detections ($9<${\sl TS}$<25$). 
{\sl TS} values of each point is shown by numbers.
Also shown by red lines are the preliminary light curves 
calculated by SPH simulation of interaction between the pulsar 
and the Be star assuming realistic parameters of the system 
\cite{Tak11} scaled arbitrary.} \label{fig:f5}
\end{figure}

Light curves in 30-day bins and 5-day bins (assuming $E^{-2}$ spectrum) are 
calculated with the help of {\sl Tools} as shown in Figures \ref{fig:f4}
(30-day bin) and \ref{fig:f5} (5-day bin).
We detected gamma-ray signal between 30 days and 65 days 
after the periastron, although there is a hint of emission 
20 days and 5 days before the periastron.
We can compare the light curve with a predicted curve
shown in red lines in Figure \ref{fig:f5} (arbitrary scaled)
which is calculated by SPH simulation of interaction
between the pulsar and the Be star \cite{Tak11}.
The general tendency of the observed light curve is not inconsitent
with prediction.

\section{DISCUSSION}

We found a significant GeV gamma-ray signal from this system 
between 30 days and 65 days after the periastron. 
Emission in this epoch should be related to the time-varying 
geometry of this system. 
We will compare the gamma-ray light curves and spectra with 
simulation \cite{Tak11} to understand the emission from this binary system.

After the analysis presented here has been completed, we found 
similar results from 
{\sl Fermi}-LAT data have been reported \cite{Tam11, Abd11}.

\bigskip 
\begin{acknowledgments}
This work is supported in part by the Grant-in-Aid 
from the Ministry of Education, Culture,
Sports, Science and Technology (MEXT)
of Japan, No. 22540315.

\end{acknowledgments}

\bigskip 

\end{document}